\documentclass[12pt]{reportj}
\usepackage{deluxetablej}
\usepackage{hyperref} 
\makeatletter
\def\@to{to}
\makeatother
\usepackage{times}
\usepackage{graphicx}
\usepackage{xcolor}
\usepackage{tocloft}
\usepackage{fancyheadings}
\emergencystretch=\maxdimen
\usepackage{datetime}
\newdateformat{ddmonthyyyy}{\THEDAY\ \monthname[\THEMONTH]\ \THEYEAR}

\renewcommand\thesection{\arabic{section}}
\renewcommand\thesubsection{\thesection.\arabic{subsection}}

\def\ssection#1{\setcounter{subsection}{0} \refstepcounter{section} \section*{\hbox to \hsize{\large\bf \arabic{section}. #1\hfill }}\label{sec} \addcontentsline{toc}{section}{\arabic{section}. #1}}
\def\ssubsection#1{\setcounter{subsubsection}{0} \refstepcounter{subsection}\subsection*{\hbox to \hsize{\normalsize\bfseries\itshape \arabic{section}.\arabic{subsection} #1\hfill}}\label{subsec} \addcontentsline{toc}{subsection}{\arabic{section}.\arabic{subsection} #1}}
\def\ssubsubsection#1{\refstepcounter{subsubsection}\subsection*{\hbox to \hsize{\normalsize\it \arabic{section}.\arabic{subsection}.\arabic{subsubsection} #1\hfill}}\label{subsubsec} \addcontentsline{toc}{subsubsection}{\arabic{section}.\arabic{subsection}.\arabic{subsubsection} #1}}

\def\ssectionstar#1{\section*{\hbox to \hsize{\large\bf #1\hfill}} \addcontentsline{toc}{section}{#1}}
\def\ssubsectionstar#1{\subsection*{\hbox to \hsize{\normalsize\bfseries\itshape #1\hfill}} \addcontentsline{toc}{subsection}{#1}}
\def\ssubsubsectionstar#1{\subsection*{\hbox to \hsize{\normalsize\it  #1\hfill}} \addcontentsline{toc}{subsection}{#1}}

\renewcommand{\cftaftertoctitle}{%
\mbox{}\hfill{\normalfont Page}}
\begin{document}

~\\

\vspace{-2.4cm}
\noindent\includegraphics*[width=0.295\linewidth]{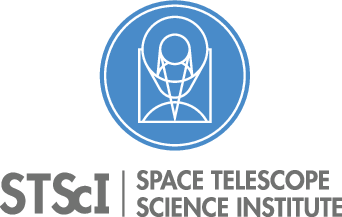}

\vspace{-0.4cm}

\begin{flushright}
    {\bf Instrument Science Report STIS 2025-04}
    
    \vspace{1.1cm}
    
    {\bf\Huge Uncertainties in Low-Count STIS Spectra}
    
    \rule{0.25\linewidth}{0.5pt}
    
    \vspace{0.5cm}
    
    Joshua D. Lothringer$^1$, Leonardo dos Santos$^1$, Joleen Carlberg$^1$, Sean Lockwood$^1$, and Jacqueline Brown$^1$
    \linebreak
    \newline
    \footnotesize{$^1$ Space Telescope Science Institute, Baltimore, MD\\}
    
    \vspace{0.5cm}
    
     \ddmonthyyyy{25 August 2025} 
\end{flushright}

\vspace{0.1cm}

\noindent\rule{\linewidth}{1.0pt}
\noindent{\bf A{\footnotesize BSTRACT}}

{\it \noindent We evaluate uncertainty calculations in the \texttt{calstis} pipeline for data in the low count regime. Due to the low dark rate and read-noise free nature of MAMA detectors, observations of UV-dim sources can result in exposures with 0 or 1 counts in some pixels. In this regime, the ``root-N" approximation widely used to calculate uncertainties breaks down, and one must compute Poisson confidence intervals for more accurate uncertainty calculations. The \texttt{CalCOS} pipeline was updated in 2020 to account for these low-count uncertainties. Here, we assess how STIS observations are currently affected by this phenomenon, describe a new Jupyter notebook exploring the issue, and introduce a new utility, \texttt{stistools.poisson\_err}, to manually calculate Poisson confidence intervals for 1D STIS spectra. Additionally, we describe a related software bug in the \texttt{stistools.inttag} utility, which splits TIME-TAG data into sub-exposures. This newly fixed bug serves as a useful case-study for the proper use of Poisson confidence intervals.}

\vspace{-0.1cm}
\noindent\rule{\linewidth}{1.0pt}

\renewcommand{\cftaftertoctitle}{\thispagestyle{fancy}}
\tableofcontents



\vspace{-0.3cm}
\ssection{Introduction}\label{sec:Introduction} 

For most astronomical measurements, observers use a ``root-N" Poisson approximation to calculate the errors or uncertainties in their data. In this approximation, the signal-to-noise is given as the square-root of the total number of events counted by the detector. However, this approximation is only valid in the high-count regime (i.e., $N\gg 10$ counts). Most astronomical observations using optical and IR detectors like charged-coupled devices (CCDs) have high enough read noise and dark current that even short exposures and exposures of dim targets result in high enough counts that the root-N approximation is appropriate.

The Space Telescope Imaging Spectrograph (STIS) instrument uses two Multi-Anode Microchannel (MAMA) detectors to observe FUV and NUV wavelengths. MAMA detectors are event-counting devices that can record individual photons at very high time-resolution (125~$\mu$s) in \texttt{TIME-TAG} mode. Two of the MAMA detectors' great advantages are very low dark current and negligible read noise. The dark rate for the STIS NUV-MAMA is around 0.001 count/sec/pix, while the dark rate for the FUV-MAMA is even lower at 0.0002 count/sec/pix. With essentially zero read noise, this means that observations of UV-dim objects (e.g., M-dwarf UV continua) can approach zero counts, even for orbit-long exposures.

The COS instrument also uses an NUV-MAMA detector identical to STIS, as well as a windowless cross delay line (XDL) in the FUV with similar noise properties. The COS team implemented a more accurate uncertainty calculation with \texttt{CalCOS} version 3.3.10, accounting for the breakdown of the root-N approximation by using a numerical approximation to Poisson confidence intervals (COS ISR 2021-03). Here, we explore the same problem as it applies to STIS data, as well as our recommended solution to observers using data in this low-count regime.

\lhead{}
\rhead{}
\cfoot{\rm {\hspace{-1.9cm} Instrument Science Report STIS 2025-04(v1) Page \thepage}}

\vspace{-0.3cm}

\ssection{Confidence Intervals}\label{sec:stats:CI}

Here, we are concerned with calculating our uncertainty on $\lambda$, the true number of counts, given some measurement of counts $N$, in the presence of random statistical noise that follows some probability distribution. We can define a confidence interval at a certain confidence level, $CL$, to quantify our uncertainty in our inference of the population mean $\lambda$. For example, a 68\% confidence level interval would correspond to the range of $\lambda$, usually a defined by an upper- and lower-limit, $\lambda_u$ and $\lambda_l$, respectively, around the measured value of $N$ such that the cumulative probability distribution  equals $ CL$. This would mean that if we repeated the measurement many times, $100*CL$\% of the confidence intervals we constructed would contain the true value of $\lambda$. This is what we mean by the uncertainty on a measured value in a frequentist statistical interpretation. 

For a Gaussian distribution, we usually quote a two-sided $CL$ interval, due in part to the distribution's symmetry about the  mean. For example, a two-sided 68\% confidence interval corresponds to a $\pm1\sigma$ range above and below the inferred population value. Often times, a ``1-$\sigma$ confidence interval" is used to mean a two-sided 68\% confidence interval. One can also quote 2-$\sigma$ confidence level at approximately 95\%, 3-$\sigma$ intervals at approximately 99.5\%, etc.

One can also quote a one-sided confidence interval, providing an upper or lower limit to our estimate of $\lambda$. This can be useful in the case of asymmetric probability distributions, where the upper and lower limits on $\lambda$ may have different functional forms or approximations, as will be the case for the Poisson distribution. In this case, a $CL$ of 0.8413 would correspond to an upper-limit on $\lambda$ such that the cumulative probability distribution equals 0.8413. This is equivalent to a two-sided ``1-$\sigma$ confidence interval" because if we subtract off the lower end's tail ($1-CL = 0.1587$), we get 0.6827, the same as the  1-$\sigma$ two-sided confidence interval. The lower limit on $\lambda$ would then correspond to this lower end tail, where the cumulative probability distribution equals $1-CL$ for a given $N$.

\ssubsection{Poisson Confidence Intervals}

The Poisson probability distribution describes the probability distribution of observed counts, $N$, of some event with an underlying `true' value $\lambda$. Formally, the Poisson distribution is described as:

\begin{equation}\label{eq:poisson}
    f(N;\lambda) = \frac{\lambda^Ne^{-\lambda}}{N!},
\end{equation}

\noindent which gives the probability of measuring $N$ counts given an expected value of $\lambda$. The variance of this distribution is $\lambda$. Some of the important properties of the Poisson distribution are:

\begin{enumerate}
    \item As $N\rightarrow\infty$, the Poisson distribution will approach the Gaussian distribution with a mean and variance of $N$.
    \item The Poisson distribution is discrete, as one can only measure integer counts. 
    \item The Poisson distribution is never negative, i.e., one can never measure a negative number of counts.
    \item Because of Points 2-3, the Poisson distribution becomes highly asymmetric at low counts because the ``left" side of the distribution cannot be negative and ``piles up" near zero.
    \item Because of Point 4, the Poisson distribution is \emph{not} well-approximated by the Gaussian distribution as $N\rightarrow 0$.
\end{enumerate}

\begin{figure}[b!]
    \centering
    \includegraphics[width=0.85\linewidth]{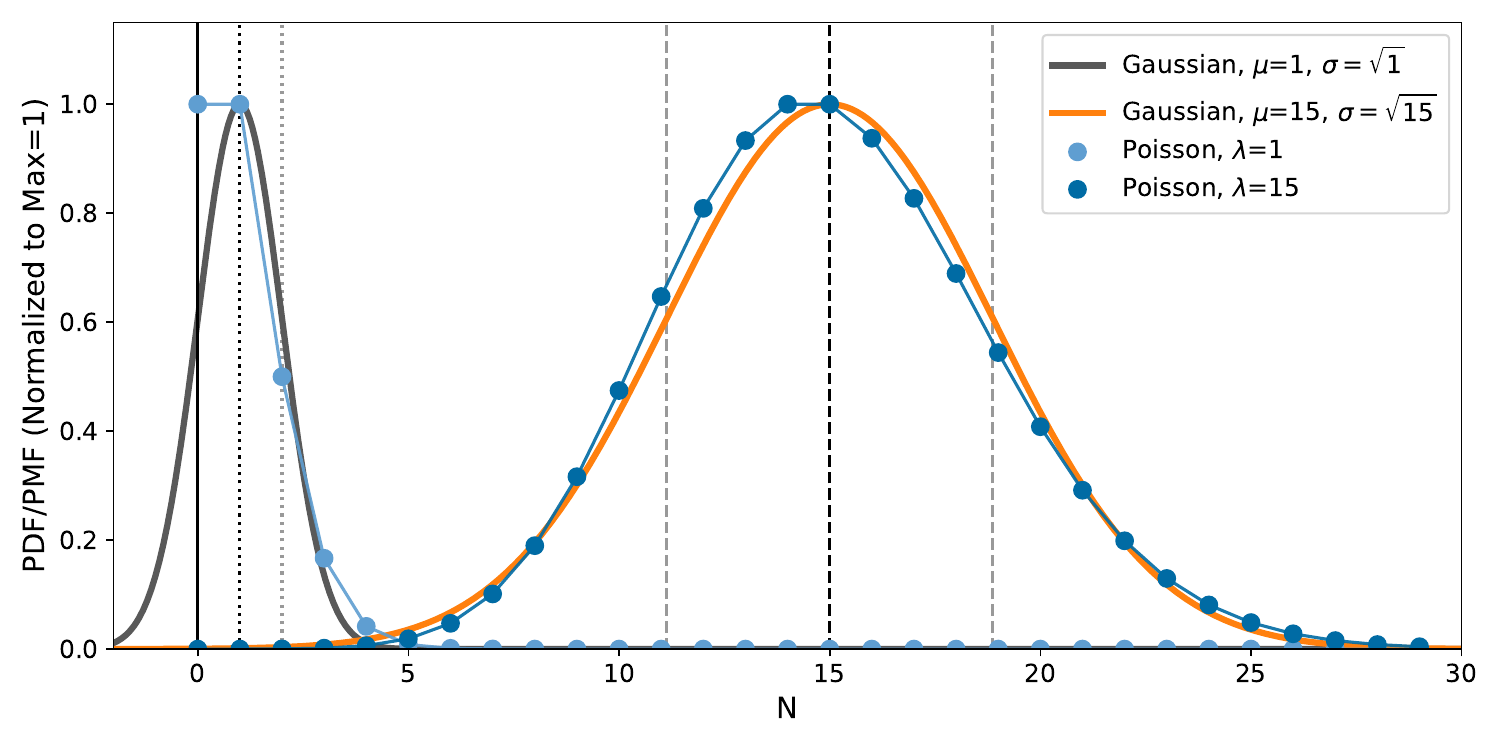}
    \caption{Poisson versus Gaussian Distribution for $\lambda = \mu = 1$ and 15. The Gaussian distributions are given $\sigma$ consistent with the root-N approximation of the Poisson distribution.}
    \label{fig:gauss_compare}
\end{figure}

\noindent Figure~\ref{fig:gauss_compare} shows how the Poisson and Gaussian distributions vary at $\lambda=1$ and $\lambda=15$. At $\lambda=1$, the difference is largest, where the Gaussian distribution allows for negative counts.

As with the Gaussian distribution, we can define confidence intervals for the Poisson distribution given some level of confidence. Unlike the Gaussian distribution though, there is no direct correspondence to the variance, nor a closed formed equation to solve for these confidence levels exactly. Additionally, the asymmetry of the Poisson distribution requires separate definitions for the upper- and lower-limits, $\lambda_u$ and $\lambda_l$, respectively, as mentioned in Section~\ref{sec:stats:CI}. Using the cumulative probability distributions, 

\begin{equation}
    \sum_{x=0}^{N-1}\frac{\lambda_l^x e^{-\lambda_l}}{x!} = CL   \hspace{25pt}   (N \neq 0) \label{eq:lower}
\end{equation}

\noindent and

\begin{equation}
    \sum_{x=0}^{N}\frac{\lambda_u^x e^{-\lambda_u}}{x!} = 1 - CL \label{eq:upper}.
\end{equation}

\noindent Because of the nature of Equations~\ref{eq:lower} and \ref{eq:upper}, one cannot simply solve for $\lambda_u$ or $\lambda_l$ at some given $CL$, which is one reason why the root-N approximation is so often relied upon. For more robust uncertainties, various numerical approximations have been developed to provide Poisson confidence intervals based on Equations~\ref{eq:lower} and \ref{eq:upper}. Within astronomy, Gehrels 1986 provides a frequentist interpretation of the Poisson confidence intervals, providing several approximations for small numbers of events/counts. One such lower confidence interval approximation is:

\begin{equation}\label{eq:ld}
    \lambda_l\approx N\left(1-\frac{1}{9N}-\frac{S}{3\sqrt{N}}\right)^3,
\end{equation}

\noindent where $N$ is the number of measured counts and $S$ is the equivalent Gaussian number of $\sigma$ corresponding to the confidence interval. Thus the lower 1$\sigma$ 15.87\% confidence interval becomes

\begin{equation}\label{eq:ld_1sig}
    \lambda_{l,1\sigma}\approx N\left(1-\frac{1}{9N}-\frac{1}{3\sqrt{N}}\right)^3.
\end{equation}

The upper confidence interval can similarly be approximated by

\begin{equation}\label{eq:lu}
    \lambda_u\approx N+S\sqrt{N+\frac{3}{4}}+\frac{S^2+3}{4},
\end{equation}

\noindent and

\begin{equation}\label{eq:lu_1sig}
    \lambda_{u,1\sigma}\approx N+\sqrt{N+\frac{3}{4}}+1,
\end{equation}

\noindent At high $N$, the difference between $N$ and both Equations~\ref{eq:ld_1sig} and \ref{eq:lu_1sig} will approach $\sqrt{N}$, hence the root-N approximation. The upper limit will always be at least about 1 higher than the root-N approximation due to the discrete nature of the Poisson confidence interval: the limit must be defined about 1 count higher to ensure that the true population mean is included in the interval at the given confidence. In the scenario where no counts are detected ($n=0$), $\lambda_{l,1\sigma}=0$ since one cannot measure negative counts, but $\lambda_{u,1\sigma} = 1.866$.

\begin{figure}
    \centering
    \includegraphics[width=0.8\linewidth]{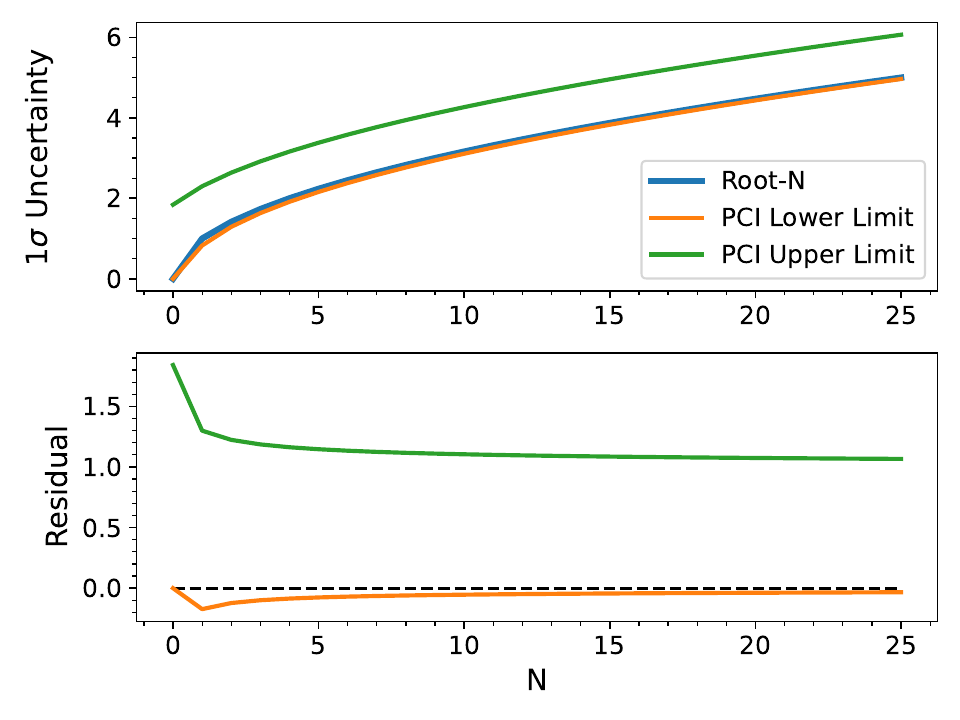}
    \caption{Top: The upper and lower limits to the 1-$\sigma$ (1-sided 84.13\%) confidence interval for the ``frequentist-confidence" interval from \texttt{astropy.poisson\_conf\_interval} compared to the root-N approximation. The lower limit from the Poisson confidence interval (PCI) mostly overlaps the root-N approximation at this scale. Bottom: The residual between the Poisson confidence intervals and the root-N approximation. The upper-limit asymptotes to always be about +1 larger than the root-N approximation due to the discrete nature of the Poisson distribution.}
    \label{fig:error_compare}
\end{figure}

A match to the Gehrels 1986 approximations is implemented in \texttt{astropy.stats} inside the \texttt{poisson\_conf\_interval}\footnote{\url{https://docs.astropy.org/en/stable/api/astropy.stats.poisson_conf_interval.html\#astropy.stats.poisson_conf_interval}} module under the ``frequentist-confidence" interval, which uses a relationship between Poisson sums and the $\chi^2$ distribution. This function is implemented in the \texttt{CalCOS} pipeline to measure their uncertainties given the total detector counts (COS ISR 2021-03). We show how the upper- and lower-limits of the ``frequentist-confidence" interval compare to the Gaussian root-N approximation in Figure~\ref{fig:error_compare}.

An alternative implementation of Poisson confidence intervals comes from Kraft, Burrows, and Nousek 1991, which provides approximations based on a Bayesian interpretation of the confidence intervals. While the classical frequentist interpretation of the confidence interval would be that it describes the probability of a repeated measurement having a confidence interval that contains the true population mean some percent of the time, the Bayesian interpretation of the confidence interval reflects the probability that sources of different flux could produce the measured count. These interpretations lead to slightly different (but more complicated) approaches in defining $\lambda_u$ and $\lambda_l$. 

Table~\ref{tab:astropyPCI} lists the upper- and lower-limits on $\lambda$ given different counts $N$ for the various confidence-interval definitions. The \texttt{root-N-0} interval is the same as the conventional \texttt{root-N} approximation, except that at $N=0$, an upper limit of 1 is given. The relatively simple \texttt{Pearson} approximation was suggested by the Collider Detector at Fermilab (CDF) group and has a few advantages over the root-N approximation, including asymmetry around $N=0$.\footnote{\url{https://web.archive.org/web/20210222093249/https://www-cdf.fnal.gov/physics/statistics/notes/pois_eb.txt}} The \texttt{sherpa-gehrels} interval comes from the Sherpa fitting package,\footnote{\url{https://cxc.cfa.harvard.edu/sherpa/statistics/}} yet, despite its name, does not correspond to any approximation in Gehrels 1986, with its lack of asymmetry resulting in negative lower limits and a general over-estimation of uncertainties. 

The last two, \texttt{frequentist-confidence} and \texttt{kraft-burrows-nousek} correspond to approximations in Gehrels 1986 and Kraft, Burrows, and Nousek 1991, respectively. While they agree well at low $N$, they begin to diverge slightly at higher $N$. The \texttt{kraft-burrows-nousek} interval is more numerically intensive and approximations for $N\geq100$ require the \texttt{mpmath} software package. Because of these considerations, the COS team at STScI chose to implement the \texttt{frequentist-confidence} confidence interval in the \texttt{CalCOS} pipeline, which we will also use for STIS data.

\begin{table}[h]
\centering
\caption{Poisson confidence intervals for different methods and observed counts using \texttt{astropy.stats.poisson\_conf\_interval()}\label{tab:astropyPCI}}
\vspace{10pt}
\begin{tabular}{lcccc}
\hline
\textbf{Method} & \multicolumn{4}{c}{\textbf{N}} \\
\cline{2-5}
 & \textbf{0} & \textbf{1} & \textbf{10} & \textbf{50} \\
\hline
\textbf{Root-N} \\
\quad $\lambda_l$ & 0.00 & 0.00 & 6.84 & 42.93 \\
\quad $\lambda_u$ & 0.00 & 2.00 & 13.16 & 57.07 \\
\textbf{Root-N-0} \\
\quad $\lambda_l$ & 0.00 & 0.00 & 6.84 & 42.93 \\
\quad $\lambda_u$ & 1.00 & 2.00 & 13.16 & 57.07 \\
\textbf{Pearson} \\
\quad $\lambda_l$ & 0.00 & 0.38 & 7.30 & 43.41 \\
\quad $\lambda_u$ & 1.00 & 2.62 & 13.70 & 57.59 \\
\textbf{Sherpa-Gehrels} \\
\quad $\lambda_l$ & -1.87 & -1.32 & 5.72 & 41.88 \\
\quad $\lambda_u$ & 1.87 & 3.32 & 14.28 & 58.12 \\
\textbf{Frequentist-Confidence} \\
\quad $\lambda_l$ & 0.00 & 0.17 & 6.89 & 42.95 \\
\quad $\lambda_u$ & 1.84 & 3.30 & 14.27 & 58.12 \\
\textbf{Kraft-Burrows-Nousek} \\
\quad $\lambda_l$ & 0.00 & 0.13 & 6.15 & 40.67 \\
\quad $\lambda_u$ & 1.84 & 3.37 & 15.19 & 60.66 \\
\hline
\end{tabular}
\end{table}

\ssection{Application to STIS Data: Current Pipeline Behavior}\label{sec:pipeline}

Within the STIS pipeline, the root-N approximation is implemented as:

\begin{equation}
    \sigma = \sqrt{\frac{N-B}{g} + \left(\frac{R}{g}\right)^2} = \sqrt{\frac{N-B}{g}} + \frac{R}{g},
\end{equation}

\noindent where $N$ is the measured number of counts (DN) on a pixel. For CCD observations, $B$ is the bias level in DN, $g$ is the gain in e$^-$/DN, and $R$ is the readnoise in e$^-$ as read from the CCD parameters table (CCDTAB) (ISR STIS 98-26). However, for MAMA observations, $B$ and $R$ are zero and dark levels are very relatively low.

\begin{figure}
    \centering
    \includegraphics[width=0.7\linewidth]{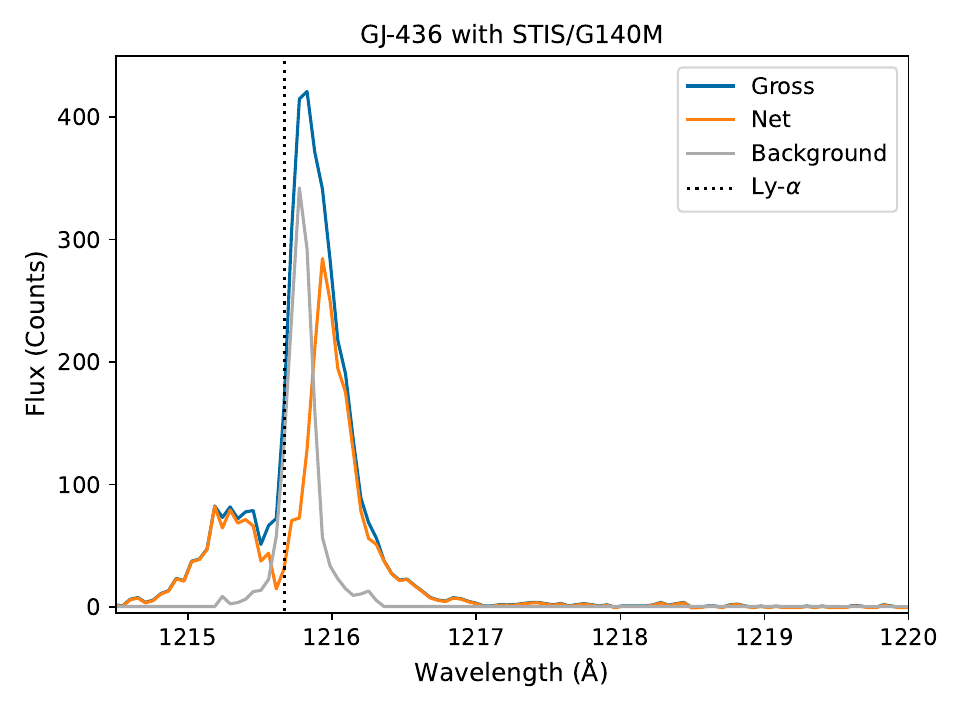}
    \includegraphics[width=0.7\linewidth]{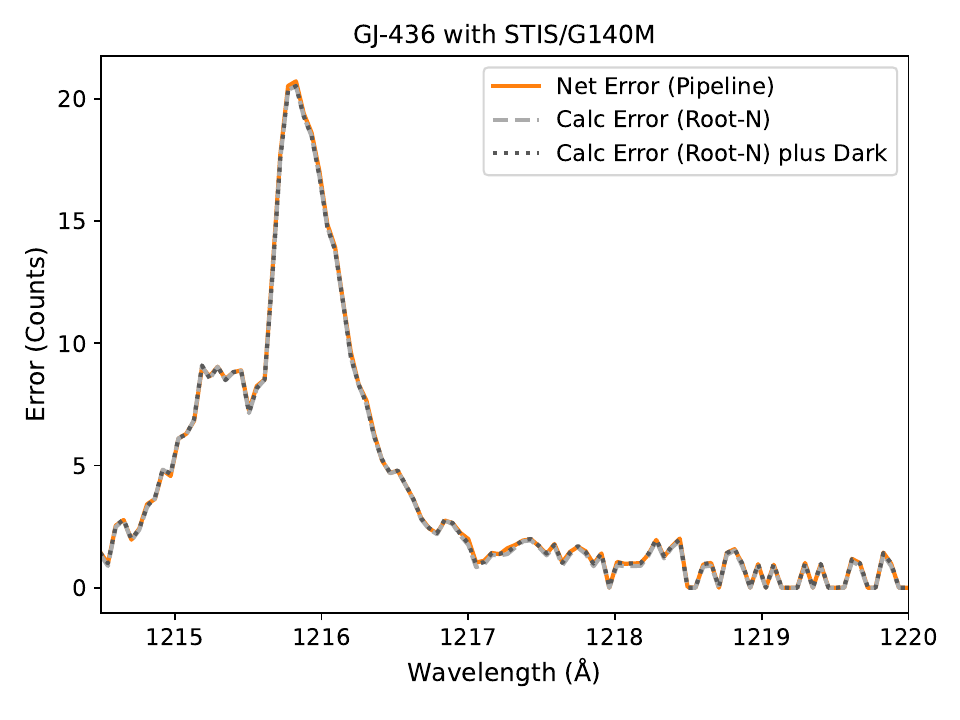}
    \caption{Top: The 1D extracted spectrum from STIS/G140M observations of M-dwarf GJ 436 just before a transit of its Neptune-sized exoplanet GJ 436b from Program 12034 (PI: Green, Dataset ID: \texttt{obgh07020}). Note the large Ly-$\alpha$ emission surrounded by regions of near-zero counts. The large background flux comes from geocoronal ``airglow" of Ly-$\alpha$, while ISM absorption of the target flux results in absorption of the stellar line's core. Bottom: Uncertainties in the spectrum from the pipeline (solid gold), the manual root-N approximation (dashed light grey), and the manual root-N approximation plus the average dark count (dotted dark grey).}
    \label{fig:gj436b}
\end{figure}

We first verified the behavior of the \texttt{calstis} pipeline by comparing pipeline uncertainties to a manual root-N approximation using HST/STIS/G140M data from Program 12034 (PI: Green, Dataset ID \texttt{obgh07020}) used to study hydrogen gas escaping from the Neptune-sized exoplanet GJ 436b (Kulow et al. 2014, Ehrenreich et al. 2015). Figure~\ref{fig:gj436b} shows Ly-$\alpha$ emission from the chromospherically-active host star, GJ 436. Interstellar hydrogen absorbs the core of the Ly-$\alpha$ emission line at about 1215.67~${\rm \AA}$, but significant emission from GJ 436 is detected in the wings of the line, even after removing the contribution from geocoronal Ly-$\alpha$ emission in the background. Outside of the Ly-$\alpha$ feature though, nearly no counts are detected from the target or background, providing a good test of our uncertainty calculations at both high- and low-$N$.

Figure~\ref{fig:gj436b} also shows the resulting uncertainty on the 1D extracted spectrum in counts from both the pipeline and a manual root-N approximation. To derive the observed number of counts, $N$, we must include counts from the target, background, and dark rate. We can calculate the counts from the target and background by multiplying the \texttt{GROSS} counts rate by the exposure time from the \texttt{EXPTIME} keyword in the data header.\footnote{We do not use the \texttt{NET} count rate, as these do not include background counts that are subtracted during reduction} As mentioned in Section~\ref{sec:Introduction}, MAMA observations have very low dark rates and no read-noise, so even exposure times of entire HST orbits (1500-3000s) are insufficient to accumulate enough counts to guarantee the validity of the root-N approximation. However, the dark rate for NUV-MAMA observations is high enough to add a few additional counts that must be taken into account when calculating the uncertainties, especially for long exposures. The dark counts have already been subtracted in the \texttt{GROSS} counts, so to gain a first-order correction, we simply use the \texttt{MEANDARK} value in the \texttt{x1d.fits} file headers. Note that this value is counts per pixel and needs to be multiplied by the extraction window height (but not the exposure time). Overall, the pipeline error is matched by the manual root-N approximation, verifying our understanding of the pipeline's behavior. 

\begin{figure}[hb]
    \centering
    \includegraphics[width=0.85\linewidth]{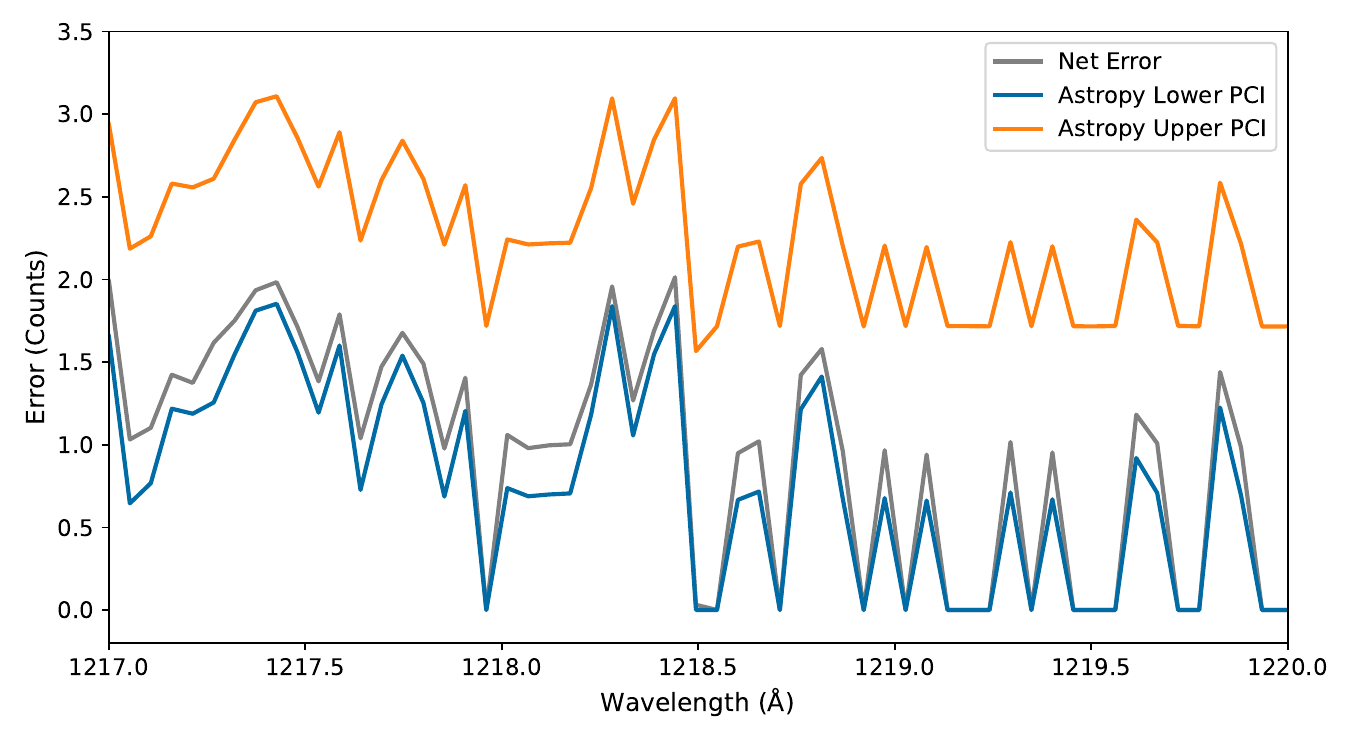}
    \caption{Uncertainties for the data show in Figure~\ref{fig:gj436b} longward of 1217~$\mathrm{\AA}$ where very few counts are detected. The pipeline-calculated uncertainties (grey) are significantly underestimated compared to the true $1\sigma$ Poisson upper 84.13\% confidence interval (orange).}
    \label{fig:gj436_PCI}
\end{figure}

While the root-N approximation is valid for the data near the Ly-$\alpha$ line, the approximation will breakdown at wavelengths longer than about 1217~$\mathrm{\AA}$, where the FUV continuum of the M-dwarf host star is not detected and counts drop to near-zero. Figure~\ref{fig:gj436_PCI} shows how the uncertainties calculated from the pipeline using the root-N approximation are underestimated compared to the proper Poisson upper confidence intervals, calculated with \texttt{astropy.stats.poisson\_conf\_interval}. The Poisson upper limit is significantly larger than the root-N pipeline calculation and, importantly, does not go to 0 when no counts are detected.

\ssection{New Tools for Calculating Poisson Errors}

\ssubsection{\texttt{stistools.poisson\_err}}\label{sec:poisson_err}

For STIS, we choose not to make direct changes to the \texttt{calstis} pipeline. Instead, we introduce a new function to the Python \texttt{stistools} package called \texttt{poisson\_err}, which can take advantage of \texttt{astropy.stats.poisson\_conf\_interval}. This tool is not automatically run as part of the default pipeline reduction but must be run manually by users in situations where the root-N approximation breaks down. Users can determine if Poisson errors should be used by running \texttt{poisson\_err} and comparing to the pipeline provided errors.

\begin{figure}[b!]
    \centering
    \includegraphics[width=0.85\linewidth]{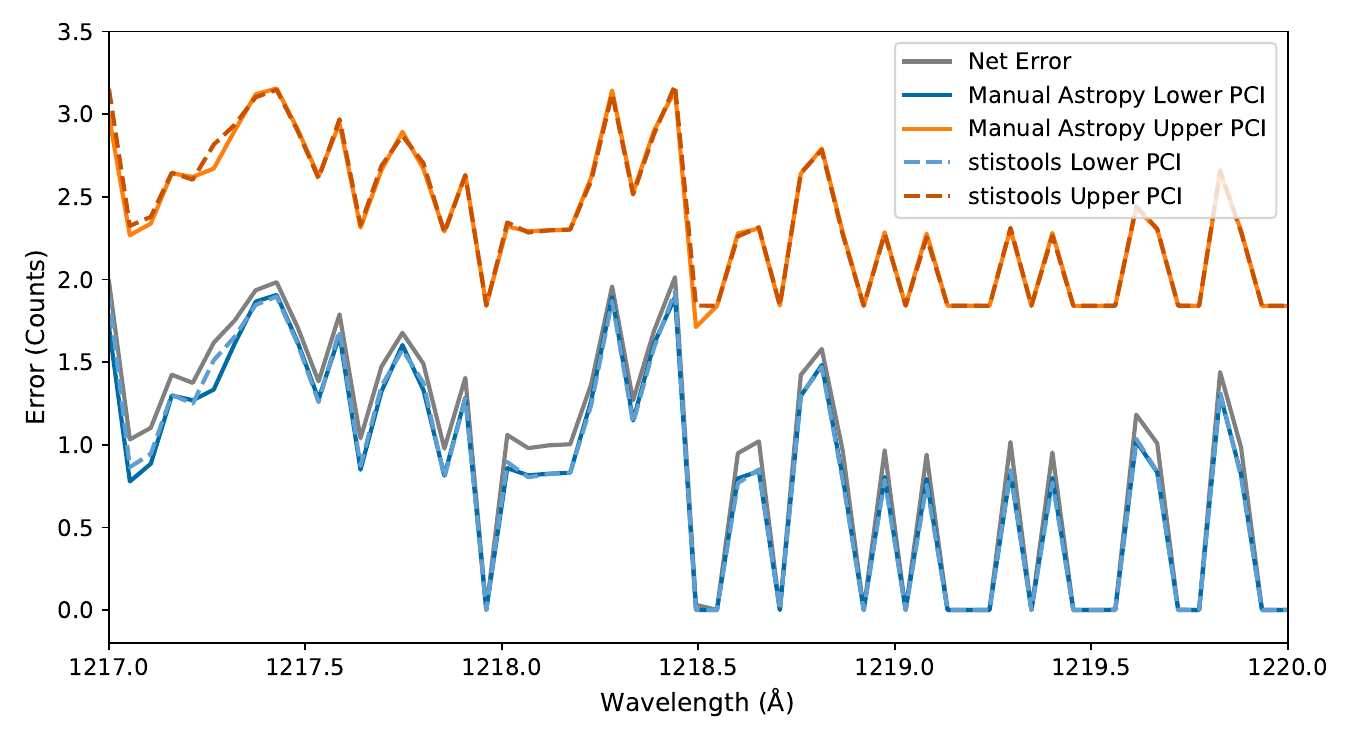}
    \caption{Demonstration of the errors calculated from \texttt{stistools.poisson\_err} using \texttt{obgh07020}. The errors from \texttt{stistools.poisson\_err} (dashed) are shown to match manual Poisson confidence intervals calculated with \texttt{astropy.stats.poisson\_conf\_interval} (solid).}
    \label{fig:poisson_err}
\end{figure}

The function can be run on any FUV or NUV-MAMA \texttt{x1d} or \texttt{sx1} file with \texttt{stistools.poisson\_err.poisson\_err(x1dfile,output)}. In Section~\ref{sec:pipeline}, we discussed the separate contribution to the error from the signal, background, and dark. Because the pipeline uses the root-N method, we can easily re-create the observed number of counts before dark subtraction by simply squaring the product of the \texttt{NET\_ERROR} rate (in counts per second) and the exposure time, \texttt{EXPTIME} (in seconds):

\begin{equation}
    N = ({\rm \texttt{NET\_ERROR}} * {\rm \texttt{EXPTIME}})^2
\end{equation}

\noindent The function then uses this observed number of counts to calculate the Poisson confidence interval using \texttt{astropy.poisson\_conf\_interval}, using the `frequentist-confidence' interval, as is done in the \texttt{CalCOS} pipeline.

The function will generate a new \texttt{x1d} or \texttt{sx1} file named \texttt{output} that will contain four new columns to each data unit called \texttt{NET\_ERROR\_PCI\_UP}, \texttt{NET\_ERROR\_PCI\_DOWN}, \texttt{ERROR\_PCI\_UP}, and \texttt{ERROR\_PCI\_DOWN} containing the 1$\sigma$ Poisson upper and lower confidence interval in both count-rate and flux units, respectively. Figure~\ref{fig:poisson_err} shows a demonstration of the agreement between \texttt{stistools.poisson\_err} and uncertainties manually calculated from \texttt{astropy.stats.poisson\_conf\_interval}.

\ssubsection{Jupyter Notebook \texttt{low\_count\_uncertainties}}

To aid users in understanding how Poisson uncertainties apply to UV HST/STIS observations, we created a Jupyter notebook in the STScI HST Notebook Repository\footnote{\url{https://spacetelescope.github.io/hst_notebooks/}} entitled ``\texttt{low\_count\_uncertainties}", which explores this problem with the same data used in Figures~\ref{fig:gj436b} and \ref{fig:gj436_PCI}. The notebook also provides a look at COS data and how the \texttt{CalCOS} pipeline has been updated to account for Poisson confidence intervals according to COS ISR 2021-03. Our recommendations to users in the notebook has recently been updated to describe the new tools described in Sections~\ref{sec:poisson_err} and \ref{sec:inttag}.

\ssubsection{\texttt{stistools.inttag}}\label{sec:inttag}

A software bug was recently discovered in the uncertainty calculations from the \texttt{stistools.inttag} utility, which splits time-tag exposures into smaller sub-exposures. When the function was populating the 2D error array for each raw sub-exposure, it was using \texttt{astropy.stats.poisson\_conf\_interval} to calculate a Poisson confidence interval. However, these uncertainties were greatly overestimated because one cannot calculate Poisson confidence intervals at the pixel-level and then combine them in quadrature during 1D spectral extraction as is done in the root-N approximation. This is because, in the root-N approximation, $\sqrt{\sum_i{(\sqrt{n_i})^2}} = \sqrt{\sum_i{n_i}} = \sqrt{N}$, where $n_i$ is the count measured on the $i$th pixel of a column and $N$ is the count for the entire column. 

\begin{figure}
    \centering
    \includegraphics[width=0.85\linewidth]{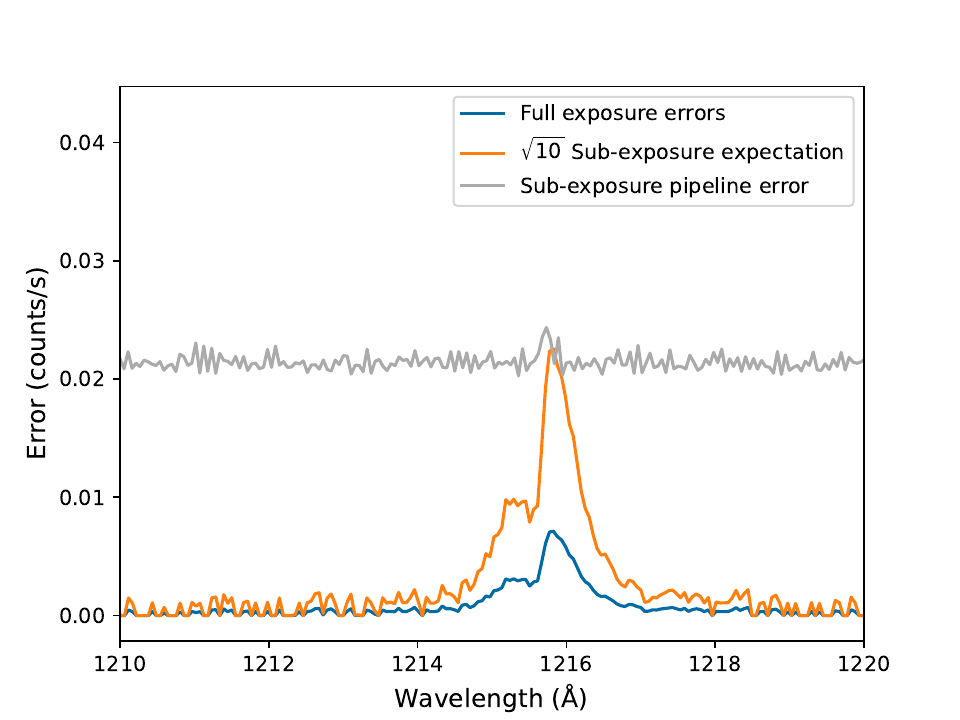}
    \caption{Comparison of the uncertainties from \texttt{obgh07020} before (blue) and after (grey) using \texttt{stistools.inttag} to split the files into 10 sub-exposures. The expected uncertainties from the root-N approximation are shown in orange, which equates to the blue line multiplied by $\sqrt{N}$.}
    \label{fig:inttag}
\end{figure}

For Poisson-confidence intervals, however, the sum in quadrature of the Poisson confidence intervals is \textit{not} equivalent to the Poisson confidence interval of the sum of the counts. As a thought-experiment, one can imagine a scenario where where no counts are detected in a hypothetical 5-pixel extraction window. The error in any one pixel would be 1.866, but adding these in quadrature gives 4.17, which is far larger than the error on the sum of counts across the window, which is is still just 1.866. The former scenario was the method used when combining and extracting \texttt{stistools.inttag} produced sub-exposures with \texttt{calstis}, which greatly over-predicted the error in sub-exposures in the low-count regime, as seen in Figure~\ref{fig:inttag}. 

In other words, it matters how the counts are distributed along a column when calculating errors like this. The \texttt{low\_count\_uncertainties} notebook provides an example, also shown in Table~\ref{tab:scenario}, of how the root-N and Poisson confidence intervals treat spectral extraction differently depending on whether 10 counts are concentrated on one pixel or evenly spread between five pixels. In each scenario, the root-N approximation gives identical values for the error, but the Poisson methods give varying estimates.

In the end, each pixel is not an independent measurement, but rather the signal from a single source spread along the cross-dispersion direction in a non-independent manner. We can therefore make the assumption that the Poisson confidence interval for the observed flux at a given wavelength ought to be calculated on the sum total of counts in a column and not at the pixel-level. Version 1.4.5 of \texttt{stistools} simply removes the Poisson confidence interval calculation from \texttt{inttag} altogether and reverts to root-N approximation. If one wants to use Poisson confidence intervals for the low-count scenarios we have discussed here, observers can use the \texttt{stistools.poisson\_err} utility described in Section~\ref{sec:poisson_err} on extracted subexposures.

\begin{table}[h]
\centering
\caption{Comparison of Error Estimates: root-N Approximation vs. Poisson Confidence Intervals}
\begin{tabular}{lcc}
\hline
\textbf{Method} & \textbf{Scenario 1} & \textbf{Scenario 2} \\
                & \textbf{[0, 0, 10, 0, 0]} & \textbf{[2, 2, 2, 2, 2]} \\
\hline
$\sqrt{N}$: add in quadrature                    & 3.16 & 3.16 \\
$\sqrt{N}$: sum then take square root            & 3.16 & 3.16 \\
Poisson: pixel-by-pixel error calculation    & 5.64 & 5.90 \\
Poisson: upper limit from total sum          & 4.27 & 4.27 \\
\hline
\end{tabular}
\label{tab:scenario}
\end{table}

\ssection{Conclusions}\label{sec:conclusion}

Here, we identified observing scenarios in which the current \texttt{calstis} pipeline uncertainty calculations breakdown, namely in very low-count regimes in the UV with MAMA detectors. As shown in Figure~\ref{fig:error_compare}, the lower uncertainty limit begins to diverge from the root-N approximation at about 20 counts. We suggest observers consider using Poisson confidence intervals at such count levels. We briefly reviewed Gaussian and Poisson statistics and explained the reason why the common root-N approximation breaks down in low-count regimes, as well as how to calculate more robust Poisson confidence intervals. 

We then described new tools to allow users to understand and calculate Poisson confidence intervals for low-count data with \texttt{stistools.poisson\_err} as demonstrated in the new \texttt{low\_count\_uncertainties} Jupyter notebook. Lastly, we identified a bug in the \texttt{stistools.inttag} routine and describe the fix that has been implemented.
   



\vspace{-0.3cm}
\ssectionstar{Acknowledgements}
\vspace{-0.3cm}

We thank Matthew Dallas for a constructive review of the initial version of this ISR.

\vspace{-0.3cm}
\ssectionstar{Change History for STIS ISR 2025-04}\label{sec:History}
\vspace{-0.3cm}
Version 1: \ddmonthyyyy{25 August 2025} - Original Document 

\vspace{-0.3cm}
\ssectionstar{References}\label{sec:References}
\vspace{-0.3cm}

\noindent
Ehrenreich, D. et al., 2015, \textit{Nature}, 522, 7557\\
Gehrels, N., 1986, ApJ, 303, 336 \\
Hodge, P., Baum, S., \& Goudfrooij, P., 1998, STIS Instrument Science Report 98-26\\
Johnson, C. et al., 2021, COS Instrument Science Report 2021-03\\
Kulow, J. et al., 2014, ApJ, 786, 2 \\

\end{document}